\documentclass[9pt]{elife}

\usepackage{lipsum} 
\usepackage[version=4]{mhchem}
\usepackage{siunitx}
\usepackage{hyperref}
\DeclareSIUnit\Molar{M}

\title{Active random force promotes diffusion in bacterial cytoplasm}

\author[1]{Lingyu Meng}
\author[2]{Yiteng Jin}
\author[3]{Yichao Guan}
\author[4]{Jiayi Xu}
\author[1,2*]{Jie Lin}
\affil[1]{Peking-Tsinghua Center for Life Sciences, Academy for Advanced Interdisciplinary Studies, Peking University, Beijing 100871, China}
\affil[2]{Center for Quantitative Biology, Academy for Advanced Interdisciplinary Studies, Peking University, Beijing 100871, China}
\affil[3]{Department of Physics, University of Illinois Urbana-Champaign, Urbana, IL 61801, USA}
\affil[4]{Yuanpei College, Peking University, Beijing 100871, China}

\corr{linjie@pku.edu.cn}{JL}

\begin{document}

\maketitle

\begin{abstract}
Experiments have found that diffusion in metabolically active cells is much faster than in dormant cells, especially for large particles. However, the mechanism of this size-dependent diffusion enhancement in living cells is still unclear. In this work, we approximate the net effect of metabolic processes as a white-noise active force and simulate a model system of bacterial cytoplasm with a highly polydisperse particle size distribution. We find that diffusion enhancement in active cells relative to dormant cells can be more substantial for large particles. Our simulations agree quantitatively with the experimental data of {\it Escherichia coli}, suggesting an autocorrelation function of the active force proportional to the cube of particle radius. We demonstrate that such a white-noise active force is equivalent to an active force of about 0.57 pN with random orientation. Our work unveils an emergent simplicity of random processes inside living cells.
\end{abstract}

\section{Introduction}

Efficient diffusion of cellular components is crucial for various biological processes in bacteria since they do not have active transport systems involving protein motors and cytoskeletal filaments \citep{von1989facilitated,elowitz1999protein,chari2010cellular,mika2011macromolecule}. Meanwhile, bacterial cytoplasm is highly crowded \citep{cayley1991characterization,zimmerman1991estimation,swaminathan1997photobleaching,ellis2001macromolecular}. Particle diffusion inside the bacterial cytoplasm is significantly suppressed compared with a dilute solution \citep{Milo2015,dix2008crowding,zhou2008macromolecular,golding2006physical,nenninger2010size,dill2011physical}. Interestingly, diffusion of large cellular components, such as plasmids, protein filaments, and storage granules, turns out to be much faster in metabolically active cells than in dormant cells, i.e., cells depleted of ATP \citep{Weber2012, Parry2014, Guo2014, Munder2016, Joyner2016}. Cellular metabolic activities appear to fluidize the cytoplasm and allow large components to sample cytoplasmic space. Another important feature of bacterial cytoplasm is its polydispersity with constituent sizes spanning from sub-nanometer to micrometers \citep{chebotareva2004biochemical,bicout1996stochastic,ando2010crowding,mcguffee2010diffusion}. Intriguingly, diffusion enhancement of a metabolically active cell relative to a dormant cell is size-dependent as large components' mobilities are much more significantly increased while small molecules diffuse virtually the same fast in active and dormant cells \citep{Parry2014}.

The physical mechanism underlying the size-dependent diffusion behaviors in active cells is far from clear. Because of the numerous ATP-consuming processes {\it in vivo}, finding the dominant biological processes that speed up diffusion may be difficult or even impossible. In a passive solution, particles receive random kicks from neighboring molecules due to thermal fluctuation. Therefore, the thermal noise's amplitude is proportional to the temperature according to the fluctuation-dissipation (FD) theorem. In contrast, in the cytoplasm of a metabolically active cell, particles also receive random kicks from biomolecules such as ATPs, amino acids, and other metabolites that do not follow detailed balance and are therefore out-of-equilibrium \citep{wilhelm2008out,gnesotto2018broken}. As a result, the net effect of their random collision with a particle is a random force not constrained by the FD theorem.

In this work, we simulate a model system of bacterial cytoplasm and adopt a coarse-grained approach by introducing an active random force as the net effect of multiple active processes in cells. We model this active random force as white noise with an amplitude independent of temperature. Our system is highly polydisperse, and according to the Stokes-Einstein relation, the autocorrelation function of the thermal random force is proportional to the particle radius. Inspired by that, we set the autocorrelation function of the active random force proportional to a power-law function of the particle radius. We find that in both passive systems without active force (corresponding to dormant cells) and active systems (corresponding to metabolically active cells), the diffusion constants are reduced under high density compared with the dilute limit. This diffusion reduction is stronger for larger particles in both passive and active systems.

Nevertheless, the enhancements of diffusion constants in active cells relative to dormant cells can be more substantial for larger particles given an appropriate size-dependent active random force. Most importantly, the experimentally measured ratios of diffusion constants between active and dormant {\it E. coli} cells agree quantitatively with our simulations and suggest an autocorrelation function of the active random force proportional to the cube of particle radius. We further demonstrate that such a white-noise active force is equivalent to an active force with a constant magnitude undergoing rotational diffusion. From the data of {\it E. coli}, we infer the magnitude of this active force as $0.57$ pN, consistent with typical force magnitudes in the cytoplasm \citep{Cusachs2017}. Our results shed light on the mechanical nature of out-of-equilibrium processes in the bacterial cytoplasm and unveil an emergent simplicity in complex living systems.

\section{Model}

\begin{figure}[htb!]
	\includegraphics[width=0.95\textwidth]{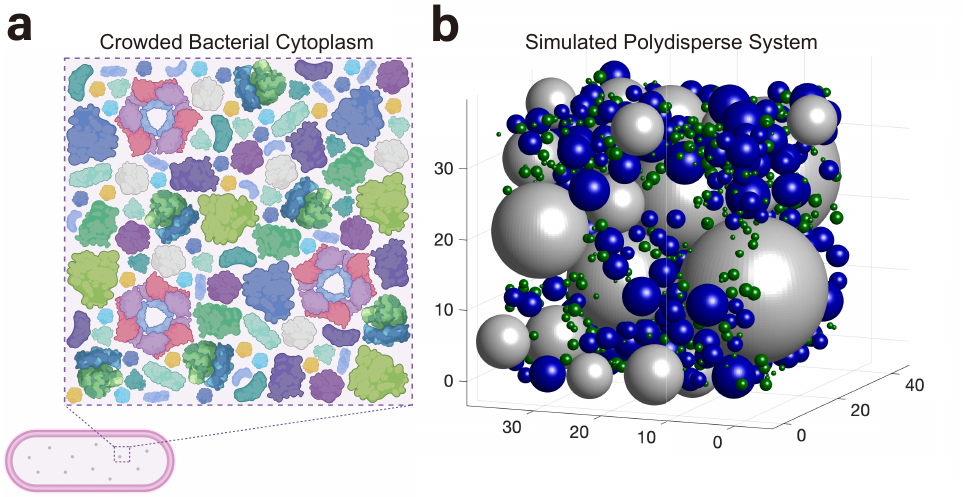}
	\caption{(\textbf{a}) A sketch of the crowded bacterial cytoplasm comprised of polydisperse cellular components. Plotted with biorender.com. (\textbf{b}) A snapshot of three-dimensional simulations using polydisperse spheres as a model system of the bacterial cytoplasm. In this example, the volume fraction $\phi=0.58$. Particles are colored according to their sizes.}
	\label{Sketch}
\end{figure}

\subsection{Size-dependent active random force}

We model the various cellular components by spherical particles with heterogeneous radii to mimic the polydisperse cytoplasm (Figure \ref{Sketch}) and their equations of motion using the Langevin dynamics,
\begin{equation}
	\eta_i \frac{\mathrm{d}r_{i,\alpha}}{\mathrm{d}t}=-\frac{\partial U}{\partial r_{i,\alpha}}+\xi_{i,\alpha}+\kappa_{i,\alpha}.  \label{rit}
\end{equation}
Here $i=1,2,..., N$ where $N$ is the number of particles and $\alpha=x,y,z$, the directions in the Cartesian coordinate. $\eta_i$ is the friction coefficient of the $i$th particle and obeys the Stokes' Law $\eta_i=6\pi\nu a_i$, where $a_i$ is the radius, and $\nu$ is the viscosity of the background solvent. $U$ is the pairwise interaction between particles which we model as $U=\frac12\sum_{i\neq j}k(a_i+a_j-r_{ij})^2\Theta(a_i+a_j-r_{ij})$ where $r_{ij}=|\mathbf{r}_i-\mathbf{r}_j|$ and $\Theta$ is the Heaviside step function: particles repel each other only when they overlap. $\xi_{i,\alpha}$ is the thermal noise, and its autocorrelation function obeys the FD theorem \citep{Doi2013}
\begin{equation}
	\langle \xi_{i,\alpha}(t) \xi_{j,\beta}(t') \rangle=12\pi \nu a_i k_BT\delta_{ij}\delta_{\alpha\beta}\delta (t-t'). 
\end{equation}
Here $k_B$ is the Boltzmann constant, and $T$ is the temperature. We introduce an active random force $\kappa_{i,\alpha}$ as the coarse-grained outcome of active processes in a metabolically active cell, and its autocorrelation function is independent of temperature,
\begin{equation}
	\langle \kappa_{i,\alpha}(t) \kappa_{j,\beta}(t') \rangle=2Aa_i^\gamma\delta_{ij}\delta_{\alpha\beta}\delta (t-t').
\end{equation}
Here we assume that the noise amplitude is a power-law function of the particle radius where $A$ and $\gamma$ are constant. Later, we will show that this size dependence of active random force is consistent with experimental measurements.

To non-dimensionalize the model, we choose the average particle radius $a_0$, the thermal energy $k_BT$, and $t_0=6\pi\nu a_0^3/k_BT$ as the length, energy, and time unit, respectively (see details in Appendix \ref{DetailsOfSimul}). Therefore, the dimensionless active noise amplitude $\tilde{A}=A a_0^{\gamma-1}/(6\pi \nu k_BT)$. In the following, variables with a tilde above are dimensionless. We simulate $N$ particles with the periodic boundary condition in a three-dimensional box. We fix the volume fraction $\phi=\sum_{i=1}^{N}\frac43\pi a_i^3/ L^3$ in each simulation, where $L$ is the side length of the system. We set the average particle radius $a_0=10$ nm. The $N$ particles' radii obey a lognormal distribution, covering two orders of magnitude, from $1$ nm to about $100$ nm, consistent with real bacterial cytoplasm \citep{Milo2015, Parry2014}. We also set $T=300$ K and choose a large spring constant $k=10^3 k_BT/a_0^2$ to mimic a hard sphere system. Details of numerical simulations are included in Appendix \ref{DetailsOfSimul}.

\section{Results}
\subsection{Active noise facilitates diffusion of large particles}

We first investigate the effects of volume fractions on the diffusion constants relative to the dilute limit for both passive and active systems. We compute the diffusion constant for each particle from the time-averaged mean square displacement (MSD) as $D= \langle \Delta \mathbf{\tilde{r}}^2 (\Delta\tilde{t})\rangle /6\Delta \tilde{t}$ with $\Delta \tilde{t}=1$ where $ \Delta \mathbf{\tilde{r}} (\Delta\tilde{t})$ is the displacement vector during a time interval $\Delta \tilde{t}$. We choose multiple values of $\phi$, including $0.58$ and $0.64$, the critical volume fractions of the glass transition, and random close packing of a monodisperse system in three dimensions \citep{Hunter2012}.

\begin{figure}[htb!]
	\includegraphics[width=0.95\textwidth]{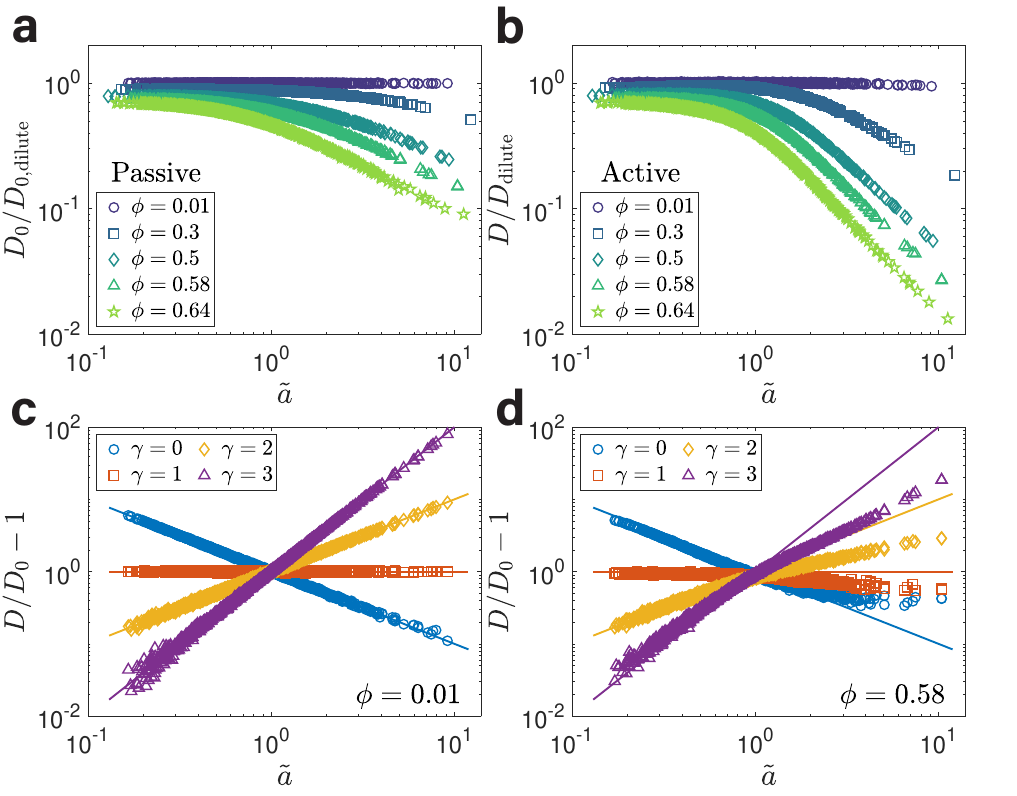}
	\caption{Diffusion constants of the simulated polydisperse systems. (\textbf{a}) The diffusion constants of passive systems relative to the dilute limit. $\tilde{a}$ is the dimensionless radius. (\textbf{b}) The same analysis as (a) but for an active system. Here $\tilde{A}=1$, $\gamma=3$. The diffusion constants in the dilute limit differ in (a) and (b). (\textbf{c}) $D/D_0-1$ for different $\gamma$'s. $D$ is the diffusion constant of the active system, and $D_0$ is the diffusion constant of the passive system. Here $\phi=0.01$, $\tilde{A}=1$. The lines with corresponding colors are the predictions in the dilute limit. (\textbf{d}) The same analysis as (c) but for systems with $\phi=0.58$. In all figures, $N=1000$.}
	\label{D_ratio}
\end{figure}

In the dilute limit, collisions between particles are negligible, and the diffusion constant of an active particle with a dimensionless radius $\tilde{a}$ becomes $D_{\text{dilute}}=1/\tilde{a} + \tilde{A} \tilde{a}^{\gamma-2}$. For a passive particle with $\tilde{A}=0$, $D_{0, \text{dilute}}=1/\tilde{a}$. In both passive and active systems, the reduction of diffusion constants in high-volume fractions relative to the dilute limit is more significant for larger particles (Figure \ref{D_ratio}a, b). To demonstrate the effects of active random force, we compare the diffusion constants of active and passive systems in the same volume fraction, which is more biologically relevant. In the dilute limit, the ratio of diffusion constants between active and dormant cells is
\begin{equation}\
	\frac{D_{\text{dilute}}}{D_{0, \text{dilute}}}=1+\tilde{A}\tilde{a}^{\gamma-1}, \label{DD0}
\end{equation}
Our simulation results for $\phi=0.01$ confirm Eq. (\ref{DD0}) (Figure \ref{D_ratio}c).

For systems with high volume fractions, neither $D$ with $\tilde{A}>0$ nor $D_0$ with $\tilde{A}=0$ is equal to the dilute limit prediction (Figure \ref{D_ratio}a, b). Intriguingly, we find that the simulation results of $D/D_0-1$ with high volume fractions still agree reasonably well with the theoretical prediction from the dilute limit, particularly for small particles with $\tilde{a}<1$ (Figure \ref{D_ratio}d). The above observation is nontrivial because all particles' diffusion constants deviate from the dilute limit regardless of size (Figure \ref{D_ratio}a, b). Deviations are observed for large particles, and we define an effective $\gamma_{\text{eff}}$ for large particles with $\tilde{a}>1.2$ such that $D/D_0-1\sim \tilde{a}^{\gamma_{\text{eff}}-1}$ (Appendix \ref{SuppleFig}---figure \ref{gamma_eff}). Experimentally, the diffusion constants of small particles are close in active and dormant cells; however, the diffusions of large particles are much faster in active cells than in dormant cells \citep{Parry2014}. Our simulations in the regime of $\gamma>1$ agree with experiments (Figure \ref{D_ratio}d). 

We also compute the diffusion constant $D=\langle \Delta \mathbf{\tilde{r}}^2 (\Delta\tilde{t})\rangle /6\Delta\tilde{t}$ using a longer $\Delta \tilde{t}$ but still short enough to ensure that the finite system size does not confine the particles' MSDs, and our conclusions are equally valid under this definition (Appendix \ref{SuppleFig}---figure \ref{D_over_Ddilute_varDt}a, \ref{D_over_Ddilute_varDt}b, and \ref{D_ratio_varDt}). We also compute the diffusion constants by fitting the MSDs using $\langle \Delta \mathbf{\tilde{r}}^2 (\Delta\tilde{t})\rangle =6D\Delta\tilde{t}$ and obtain similar results (Appendix \ref{SuppleFig}---figure \ref{D_over_Ddilute_varDt}c and \ref{D_over_Ddilute_varDt}d). 

\begin{figure}[htb!]
	\includegraphics[width=0.95\textwidth]{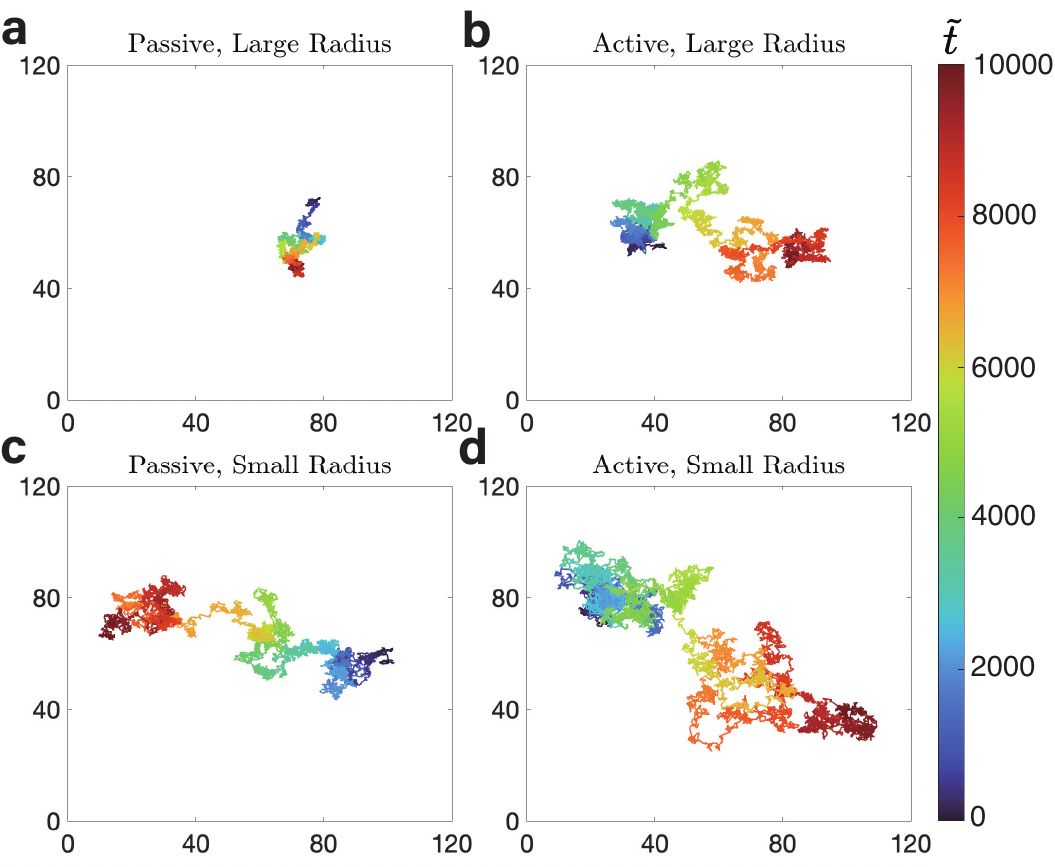}
	\caption{Single particles' trajectories projected to two dimensions. (\textbf{a}) The trajectory of a large particle with $\tilde{a}=6.75$ in a passive system. The color represents the time elapsed from the beginning of the trajectory. (\textbf{b}) The trajectory of the same particle in (a) but in an active system. (\textbf{c}) The trajectory of a small particle with $\tilde{a}=2.44$ in a passive system. (\textbf{d}) The trajectory of the same particle in (c) but in an active system. In (b) and (d), $\tilde{A}=0.42$. and $\gamma=3$. In all figures, $\phi=0.58$, $N=1000$.}
	\label{3Dpath}
\end{figure}

We track the trajectories of single particles (Figure \ref{3Dpath}). We find that particles with small radii can equally explore the system in active and passive systems; however, particles with large radii can only explore space in active systems and remain localized in passive systems, consistent with experimental observations \citep{Parry2014}. 

\begin{figure}[htb!]
	\includegraphics[width=0.8\textwidth]{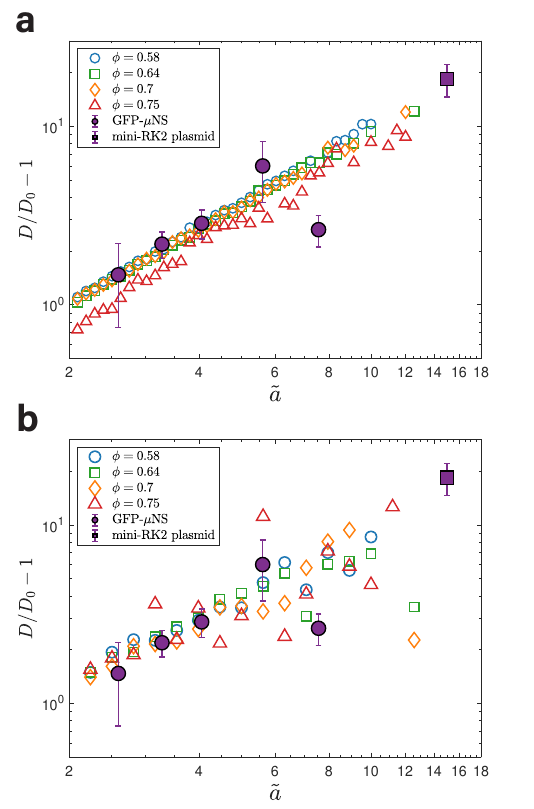}
	\caption{Comparison between simulations and experiments. (\textbf{a}) The relative enhancements of diffusion constants ($D/D_0-1$) from the experimental data match the simulation results of the white-noise active force model. Here $\gamma=3$, $\tilde{A}=0.42$. (\textbf{b}) The same analysis as (a) but for simulations of the self-propelled model (see details in Appendix \ref{DetailsOfSimul}). Here, $\tilde{F}=1.38$. The results are binned over particles with a bin interval of 0.02 in (a) and of 0.05 in (b) in the log$_{10}$ scale. In both figures, $N=4000$.}
	\label{D_ratio_wExp}
\end{figure}

\subsection{Comparison with experiments}

In \cite{Parry2014}, the authors measured the MSDs of exogenous particles of multiple sizes in active and dormant cells, including GFP-$\mu$NS particles with changeable sizes and mini-RK2 plasmid. We find that the measured MSDs are subject to large noises, making it difficult to compute the diffusion constants accurately. To circumvent this problem, we compute the ratios of MSDs in active and dormant cells at every moment and calculate the ratios of diffusion constants between active and dormant cells $D/D_0$ as the averaged MSD ratios over time. The experimental particle radius is converted to a dimensionless number using $a_0=10$ nm.

To compare with experiments, we simulate several different volume fractions since the actual volume fraction of bacterial cytoplasm is unknown. Intriguingly, the simulated $D/D_0-1$ with $\gamma=3$ nicely matches the experimental data (Figure \ref{D_ratio_wExp}a), and we will explain the physical mechanism of $\gamma=3$ later. We find that the simulated diffusion enhancements $D/D_0-1$ are insensitive to the volume fraction, although the diffusion constants $D$ and $D_0$ by themselves change significantly with the volume fraction (Figure \ref{D_ratio}a, b). To confirm the robustness of our results, we also simulate a narrower distribution of particle radii that is more Gaussian-like. The agreement between simulations and experiments is equally valid (Appendix \ref{SuppleFig}---figure \ref{SigmaLow}). Our results are also independent of the length unit as we choose a different length unit and obtain the same results (Appendix \ref{SuppleFig}---figure \ref{a0_2nm}). Using the alternative definition of diffusion constants does not affect our conclusions (Appendix \ref{SuppleFig}---figure \ref{D_ratio_wExp_varDt}).

We note that for a dilute active system with $\gamma=3$, the absolute diffusion constant is a non-monotonic function of particle size ($D_{\text{dilute}}=1/\tilde{a} + \tilde{A} \tilde{a}$). Nevertheless, we find that the absolute diffusion constant continuously decreases with particle size in systems with large volume fractions (Appendix \ref{SuppleFig}---figure \ref{D_multi_phi}), consistent with the experimental observations \citep{kumar2010mobility}. 

We also calculate the radius of gyration, the root-mean-square distance from the center of the trajectory, for both passive and active systems (Appendix \ref{SuppleFig}---figure \ref{Rg}a). The radii of gyration from simulations decrease linearly with the particle radius in both passive and active systems. The ratio between the passive and active systems also has a linear relationship with the particle radius (Appendix \ref{SuppleFig}---figure \ref{Rg}b). These results agree with experiments \citep{Parry2014}, further supporting the validity of our simulations.

\begin{figure}[htb!]
	\includegraphics[width=0.95\textwidth]{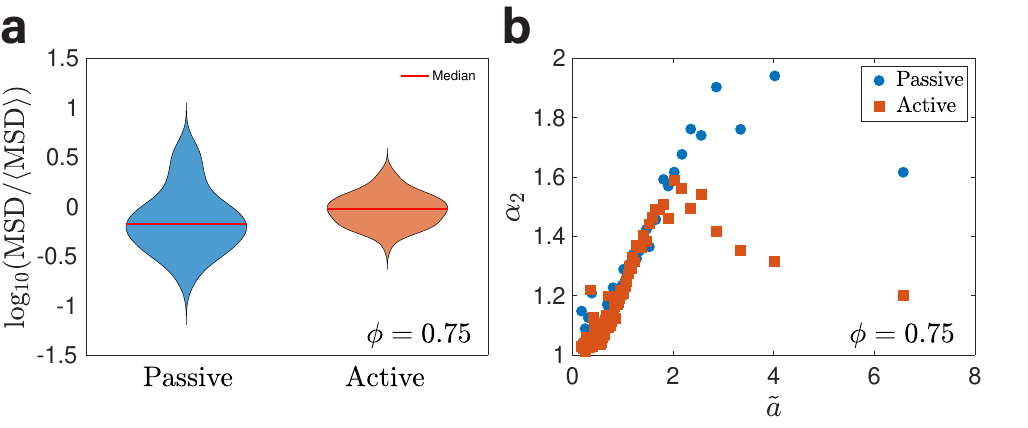}
	\caption{Activity fluidizes the glassy polydisperse system under a high volume fraction. (\textbf{a}) Violin plot of MSDs of particles whose $\tilde{a}\approx4$ for passive and active systems under $\phi=0.75$. The MSDs within a dimensionless time 1000 are shown in the log scale and normalized by the average value.  (\textbf{b}) The non-Gaussian parameter $\alpha_2$ as a function of the particle radius for passive and active systems under $\phi=0.75$. $\alpha_2=\langle \Delta \mathbf{\tilde{r}}(\Delta \tilde{t})^4 \rangle/(3\langle \Delta \mathbf{\tilde{r}}(\Delta \tilde{t})^2 \rangle^2)$, which equals $1$ if the displacements obey the Gaussian distribution. The results are averaged with 50 particles in each bin. For the active systems, $\tilde{A}=0.42$ and $\gamma=3$. In both figures, $N=4000$.}
	\label{DyHeteroAndNonGauss}
\end{figure}

In \cite{Parry2014}, the authors observed much stronger glassy-like properties in dormant cells than in living cells. We find similar hallmarks of glass transition in our simulations, including dynamic heterogeneity and non-Gaussian displacements \citep{Marcus1999, Weeks2000, Kegel2000}. The MSDs of particles with similar radii in the same passive system vary over two orders of magnitude (Figure \ref{DyHeteroAndNonGauss}a), showing significant dynamic heterogeneity. Meanwhile, this dynamic heterogeneity is much weaker in the corresponding active system. The displacement distributions have a non-Gaussian tail both in the passive and active systems (Appendix \ref{SuppleFig}---figure \ref{StepFreq}); however, the non-Gaussian degree of displacement distributions for large particles is much weaker in the active system than in the passive system (Figure \ref{DyHeteroAndNonGauss}b). Our results show that activity fluidizes the glassy polydisperse system, consistent with the experimental observations \citep{Parry2014}. We find that a higher volume fraction is needed to observe hallmarks of glass transition in polydisperse systems (Appendix \ref{SuppleFig}---figure \ref{DHNGLow}) than in monodisperse systems (Appendix \ref{SuppleFig}---figure \ref{alpha_PolyMono}), consistent with observations that polydispersity can smear out glass transition \citep{Zaccarelli2015}.

\subsection{Mechanism of the cubic scaling \boldmath{$\gamma=3$}}

In the following, we explain the cubic scaling of the white-noise active force with the particle radius. We consider a self-propelled model in which an active force with a constant magnitude is exerted on each particle \citep{Howse2007, Fily2012, Bialke2012, Redner2013}. The orientation of this active force is random due to the rotational diffusion of the particle. The equation of motion for the $i$th active particle becomes
\begin{equation}\label{NDRotMotion}
	\eta_i	\frac{\mathrm{d} r_{i,\alpha}}{\mathrm{d}t}=-\frac{\partial U}{\partial r_{i,\alpha}}+\xi_{i,\alpha}+ Fn_{i,\alpha}, 
\end{equation}
where $F$ is the magnitude of the active force and $n_{i,\alpha}$ is the orientation vector of the active force in the direction $\alpha$. The orientation vector $\mathbf{n}_i$ obeys $\mathrm{d}\mathbf{n}_i/\mathrm{d} t=\mathbf{T}_i\times \mathbf{n}_i$, where $\mathbf{T}_i$ is the thermal random torque. Its autocorrelation function satisfies the FD theorem, $\langle T_{i,\alpha}( t), T_{j,\beta}(t^{\prime})\rangle=2D_{R,i}\delta_{ij}\delta_{\alpha\beta}\delta(t-t^{\prime})$. Here $D_{R,i}=k_BT/8\pi \nu a_i^3$ is the rotational diffusion constant for a spherical particle with radius $a_i$.

At long times, the additional diffusion constant due to activity $D_{\text{active}}=(F/6\pi \nu a)^2 \times (1/2D_R)/3$, where $F/6\pi \nu a$ is the speed of the active particle and $1/2D_R$ is the time for the active force to change its orientation in three-dimensional space. Therefore, the diffusion enhancement of an active particle relative to a passive particle in the dilute limit becomes
\begin{equation}
	\frac{D_{\text{dilute}}}{D_{0, \text{dilute}}}= 1+\frac{2\tilde{F}^2}{9}\tilde{a}^2, \label{DD01}
\end{equation}
Here $\tilde{F}$ is the dimensionless active force with unit $k_B T/a_0$. Comparing Eq. (\ref{DD0}) and Eq. (\ref{DD01}), we find that the two models are equivalent in terms of diffusion enhancement when $\gamma=3$. This conclusion applies to the dilute limit, and we hypothesize that the equivalence of the two models is still valid under high-volume fractions.

To test our hypothesis, we simulate the self-propelled model with $\tilde{F}$ satisfying $\tilde{A}=2\tilde{F}^2/9$ so that the two models lead to the same diffusion enhancement in the dilute limit. The agreement between simulations and experimental data also holds for the self-propelled model (Figure \ref{D_ratio_wExp}b). We find that the magnitude of the active force $F=0.57$ pN in the physical unit, consistent with typical force magnitudes in the cytoplasm \citep{Cusachs2017}. We remark that the constant magnitude of the active force is crucial to obtain the correct scaling of diffusion enhancement. In an alternative model with a constant active speed, $D_{\text{dilute}}/D_{0, \text{dilute}}-1\sim\tilde{a}^4$, inconsistent with experiments.

\section{Discussion}

In this work, we introduce a white-noise active force to a highly polydisperse system to mimic bacterial cytoplasm. While prior works have investigated the effect of activities on particle mobility \citep{Mandal2016, Yuan2019, Abbaspour2021}, our work is the first one simultaneously incorporating crowding by different particle sizes and active forces as far as we realize. Due to its out-of-equilibrium nature, the FD theorem does not constrain the active random force. Surprisingly, a white-noise active force reproduces the experimentally measured ratios of diffusion constants between living and dormant bacteria with its autocorrelation function proportional to the cube of particle radius. We note that an active random force generally generates an additional friction coefficient \citep{Solon2022, Shakerpoor2021, Granek2022}. We argue that this friction coefficient is negligible because the active temperature is typically much higher than the thermal temperature (see details in Appendix \ref{friction}).

We further demonstrate that the white-noise active force model with $\gamma=3$ is equivalent to the self-propelled model with a constant-magnitude active force regarding the diffusion enhancement. Our results suggest an emergent simplicity when many active processes are averaged simultaneously. Importantly, we identify the magnitude of the active force, $F=0.57$ pN, which experiments can test. We note that the time-averaged MSDs of particles of the same size differ among independent simulations. Nevertheless, the MSD averaged over the time-averaged MSDs of independent simulations are close to the ensemble-averaged MSD (Appendix \ref{SuppleFig}---figure \ref{MSD_and_MSDtau}), suggesting a weak non-ergodicity effect, presumably because polydispersity smears out glass transition (Appendix \ref{SuppleFig}---figure \ref{alpha_PolyMono}). Finally, we remark that while hydrodynamic interaction has been shown to reduce diffusion coefficient, its effect may be negligible for particles with a radius above $25$ nm that we use to compare with experimental data \citep{ando2010crowding}.

\section{Acknowledgments}

We thank Yiyang Ye, Hua Tong, Sheng Mao, and Ming Han for helpful discussions related to this work. The research was funded by National Key R\&D Program of China (2021YFF1200500) and supported by grants from Peking-Tsinghua Center for Life Sciences.

\bibliography{Lin_Apr_2021}


\appendix
\begin{appendixbox}
\label{DetailsOfSimul}
\section{Details of numerical simulations}

To non-dimensionalize Eq. (1) in the main text, we choose the average particle radius $a_0$, the thermal energy $k_BT$, and $t_0=6\pi\nu a_0^3/k_BT$ as the length, energy and time unit respectively. The dimensionless equation of motion becomes
\begin{equation}\label{NDEqMotion}
	\frac{\mathrm{d}\tilde{r}_{i,\alpha}}{\mathrm{d}\tilde{t}}=-\frac{1}{\tilde{a}_i}\frac{\partial\tilde{U}}{\partial\tilde{r}_{i,\alpha}}+\tilde{\xi}_{i,\alpha}+\tilde{\kappa}_{i,\alpha}, 
\end{equation}
where $\tilde{\xi}_{i,\alpha}$ is the dimensionless thermal noise with its autocorrelation function $\langle \tilde{\xi}_{i,\alpha}(\tilde{t} ), \tilde{\xi}_{j,\beta}(\tilde{t}')\rangle=2\tilde{a}_i^{-1}\delta_{ij}\delta_{\alpha\beta}\delta(\tilde{t}-\tilde{t}')$,
and $\tilde{\kappa}_{i,\alpha}$ is the dimensionless active noise with its autocorrelation function $\langle \tilde{\kappa}_{i,\alpha}(\tilde{t} ), \tilde{\kappa}_{j,\beta}(\tilde{t}')\rangle=2\tilde{A}\tilde{a}_i^{\gamma-2}\delta_{ij}\delta_{\alpha\beta}\delta(\tilde{t}-\tilde{t}')$. Here $\tilde{A}=A a_0^{\gamma-1}/(6\pi \nu k_BT)$.

We simulate $N$ particles in a three-dimensional cubic box with a dimensionless side length $\tilde{L}$ under the periodic boundary condition with MATLAB. The particles' radii obey a shifted lognormal distribution, $\tilde{a}=\tilde{a}_{\text{min}}+\exp[\mu+\sigma \mathcal{N}(0,1)]$, where $\mathcal{N}(0,1)$ is standard normal random number, and $\mu$ and $\sigma$ are constants. The minimum radius $\tilde{a}_{\text{min}}$ is $0.1$. We set $\sigma=0.85$ to mimic the polydisperse environment and then obtain $\mu$ by setting the average radius $\tilde{a}_{\text{mean}}=\tilde{a}_{\text{min}}+\exp(\mu+\sigma^2/2)$ as $1$. Different $\phi$ are obtained by adjusting $\tilde{L}$ given $\phi=N\frac43\pi\langle\tilde{a}^3\rangle/\tilde{L}^3$, where $\langle\tilde{a}^3\rangle$ is calculated from the shifted lognormal distribution function. We choose the attempted $N$, and after $\tilde{L}$ is determined, we take $\tilde{L}$ as the nearest larger integer for numerical convenience. Therefore, the actual $N$ is slightly different from the number we set. The random radii $\tilde{a}_i$ are repeatedly generated until the actual volume fraction $\phi_{\text{real}}$, which is $\sum_{i=1}^{N}\frac43\pi\tilde{a}_i^3/\tilde{L}^3$, is close enough to the required volume fraction, $|\phi_{\text{real}}-\phi|<10^{-4}$. The total simulation time $\tilde{t}_{\text{tot}}$ is $10000$ with a time step $\mathrm{d}\tilde{t}=2\times10^{-4}$.  For all simulations, the dimensionless $\tilde{k}$ is $1000$. In Figure \ref{D_ratio_wExp}b of the main text, we use  $\Delta \tilde{t}$ where \ $\langle \Delta \mathbf{\tilde{r}}^2 (\Delta\tilde{t})\rangle = \tilde{L}^2/32$ to compute the diffusion constant so that the finite system size does not affect the results. We confirm that the $\Delta \tilde{t}$ calculated in this way is longer than the rotational relaxation time.

\end{appendixbox}

\begin{appendixbox}
\label{SuppleFig}

\section{Supplemental Figures}

\begin{center}
	\includegraphics[width=0.6\textwidth]{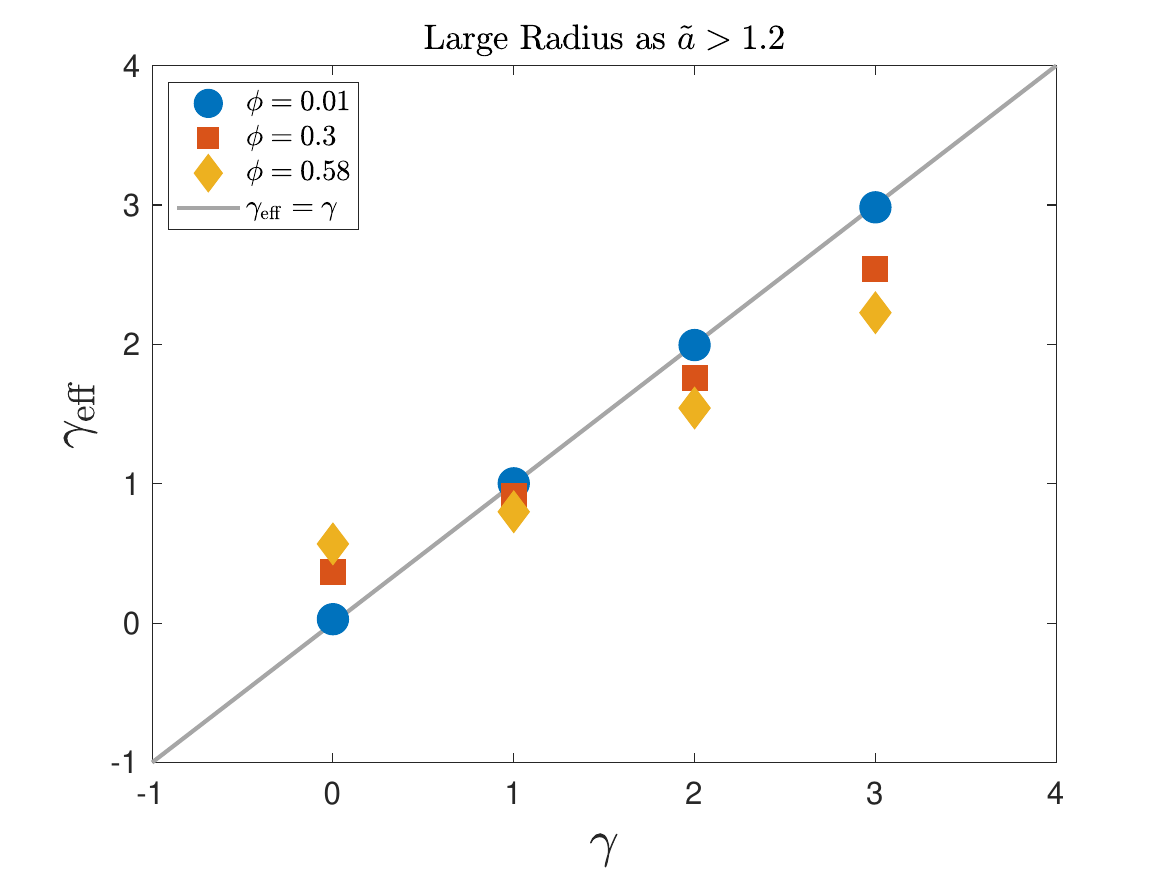}\\
	\captionof{figure}{The effective exponent $\gamma_{\text{eff}}$ {\it vs.} $\gamma$ for large particles with $\tilde{a}>1.2$. The exponent $\gamma_{\text{eff}}$ is equal to $\gamma$ in the dilute limit. Under high volume fractions, $\gamma_{\text{eff}}>\gamma$ for $\gamma<1$ and $\gamma_{\text{eff}}<\gamma$ for $\gamma>1$. Here, $N=1000$.}
	\label{gamma_eff}
\end{center}
	
\begin{center}
	\includegraphics[width=0.9\textwidth]{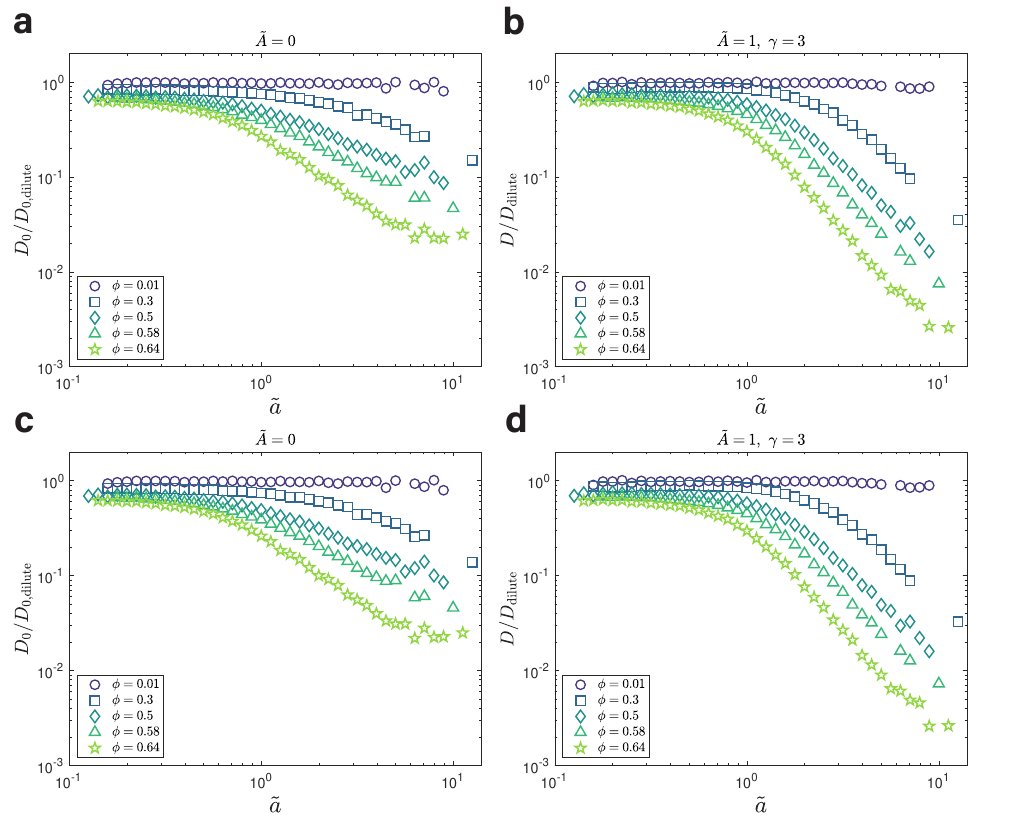}\\
	\captionof{figure}{Diffusion constants of the simulated polydisperse systems with different methods. (a) The ratios of the diffusion constants of a passive system relative to the dilute limit using a larger $\Delta \tilde{t}$. $\Delta \tilde{t}$ is the time interval to compute the diffusion constant, as $D=\langle \Delta \mathbf{\tilde{r}}^2 (\Delta\tilde{t})\rangle /6\Delta\tilde{t}$. $\tilde{a}$ is the dimensionless radius. (b) The same analysis as (a) but for an active system. (c) A similar analysis as (a) with the diffusion constants obtained by fitting the MSDs as $\langle \Delta \mathbf{\tilde{r}}^2\rangle = 6D \Delta \tilde{t}$ between $\Delta \tilde{t}/10$ and $\Delta \tilde{t}$. The results are very close to (a). (d) The same analysis as (c) but for an active system. In all figures, $\Delta \tilde{t}$ is the time where the mean square displacement $\langle \Delta \mathbf{\tilde{r}}^2 (\Delta\tilde{t})\rangle = \tilde{L}^2/32$ so that the finite system size does not affect the results. The results are binned over particles with a bin interval of $0.05$ in the log$_{10}$ scale. For the active systems, $\tilde{A}=1$ and $\gamma=3$. In all figures, $N=1000$. }
	\label{D_over_Ddilute_varDt}
\end{center}

\begin{center}
	\includegraphics[width=0.6\textwidth]{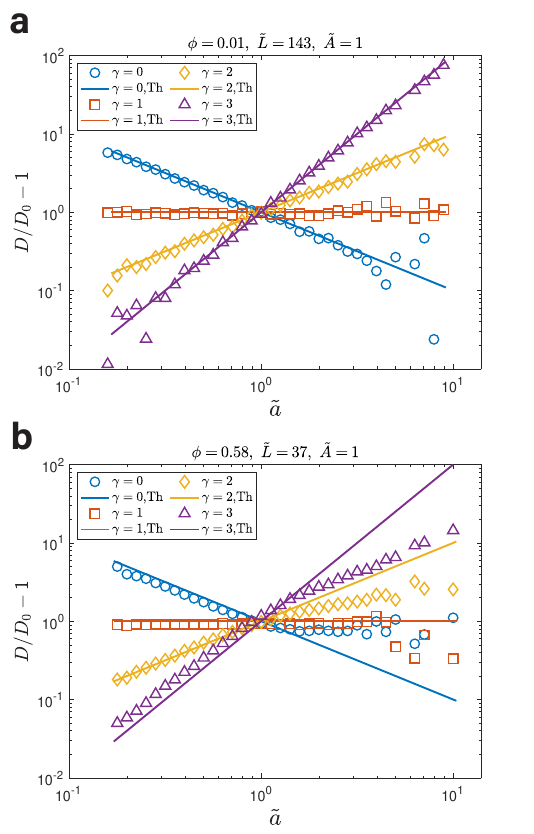}\\
	\captionof{figure}{The relative enhancement of the diffusion constants $D/D_0-1$ using a larger $\Delta \tilde{t}$ (see Appendix \ref{SuppleFig}---figure \ref{D_over_Ddilute_varDt}). (a) $\phi=0.01$ and (b) $\phi=0.58$. The results are binned over particles with a bin interval of $0.05$ in the log$_{10}$ scale. The solid lines are the theoretical prediction in the dilute limit. For the active systems, $\tilde{A}=1$. In both figures, $N=1000$. }
	\label{D_ratio_varDt}
\end{center}

\begin{center}
	\includegraphics[width=0.6\textwidth]{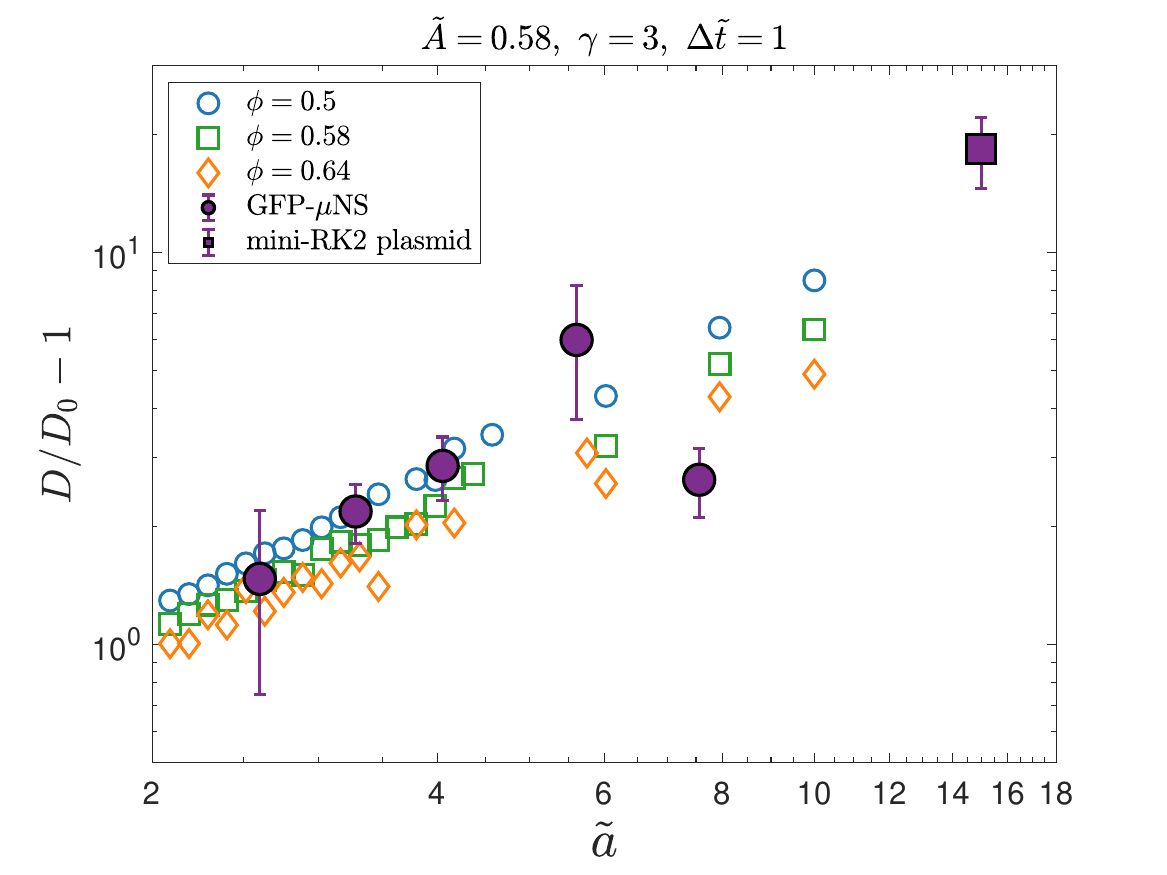}\\
	\captionof{figure}{Comparison between simulations and experiments using the lognormal distribution with $\sigma=0.5$. Three particles with radii equal to $6$, $8$ and $10$ are added to the system to mimic the experimental condition and cover a more extensive range of radius. The relative enhancements of diffusion constants from the experimental data match the simulation results reasonably well. Here, $\tilde{A}=0.58$ and $\gamma=3$. The results are binned over particles with a bin interval of $0.02$ in the log$_{10}$ scale. In this figure, $N=4000$. }
	\label{SigmaLow}
\end{center}

\begin{center}
	\includegraphics[width=0.6\textwidth]{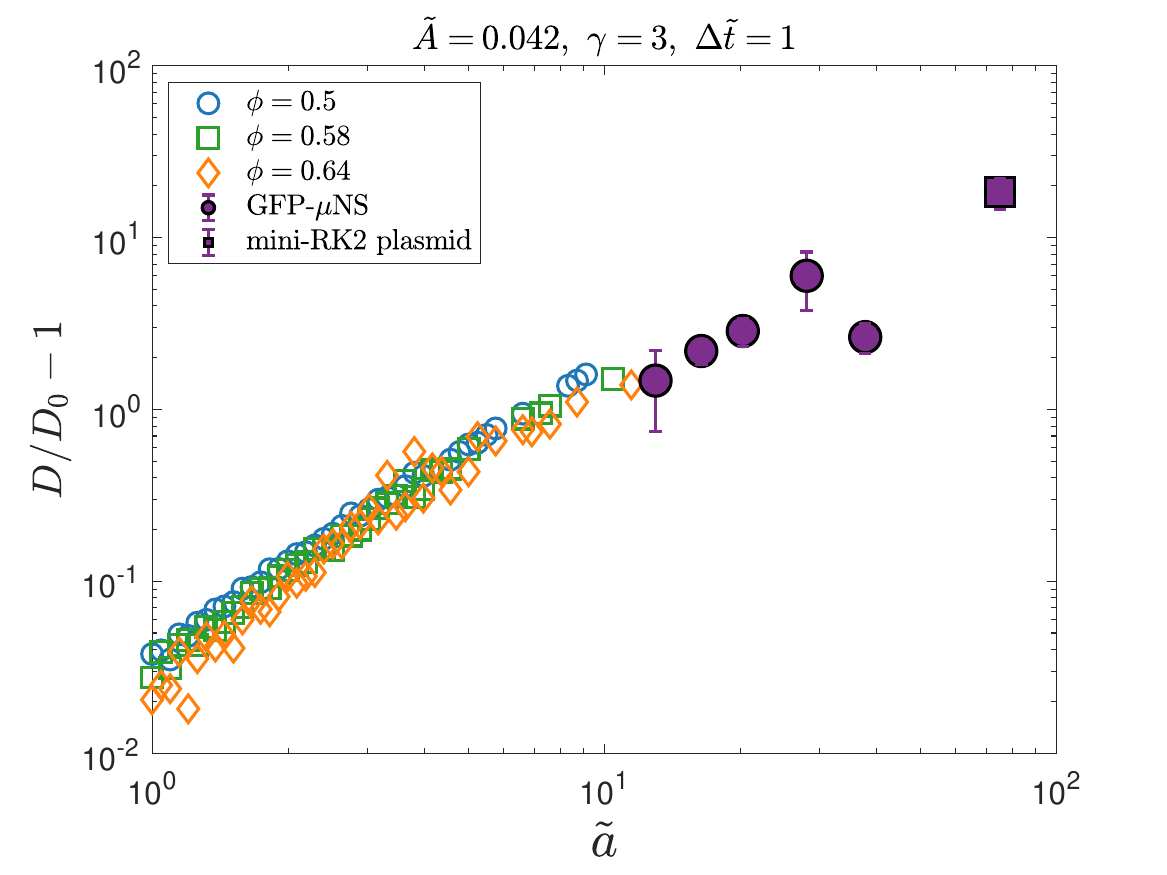}\\
	\captionof{figure}{Comparison between simulations and experiments with $a_0=2$ nm. The relative enhancements of diffusion constants ($D/D_0-1$) from the experimental data match the simulation results reasonably well. Here, $\tilde{A}=0.042$ and $\gamma=3$. The results are binned over particles with a bin interval of $0.02$ in the log$_{10}$ scale. In this figure, $N=1000$. }
	\label{a0_2nm}
\end{center}

\begin{center}
	\includegraphics[width=0.6\textwidth]{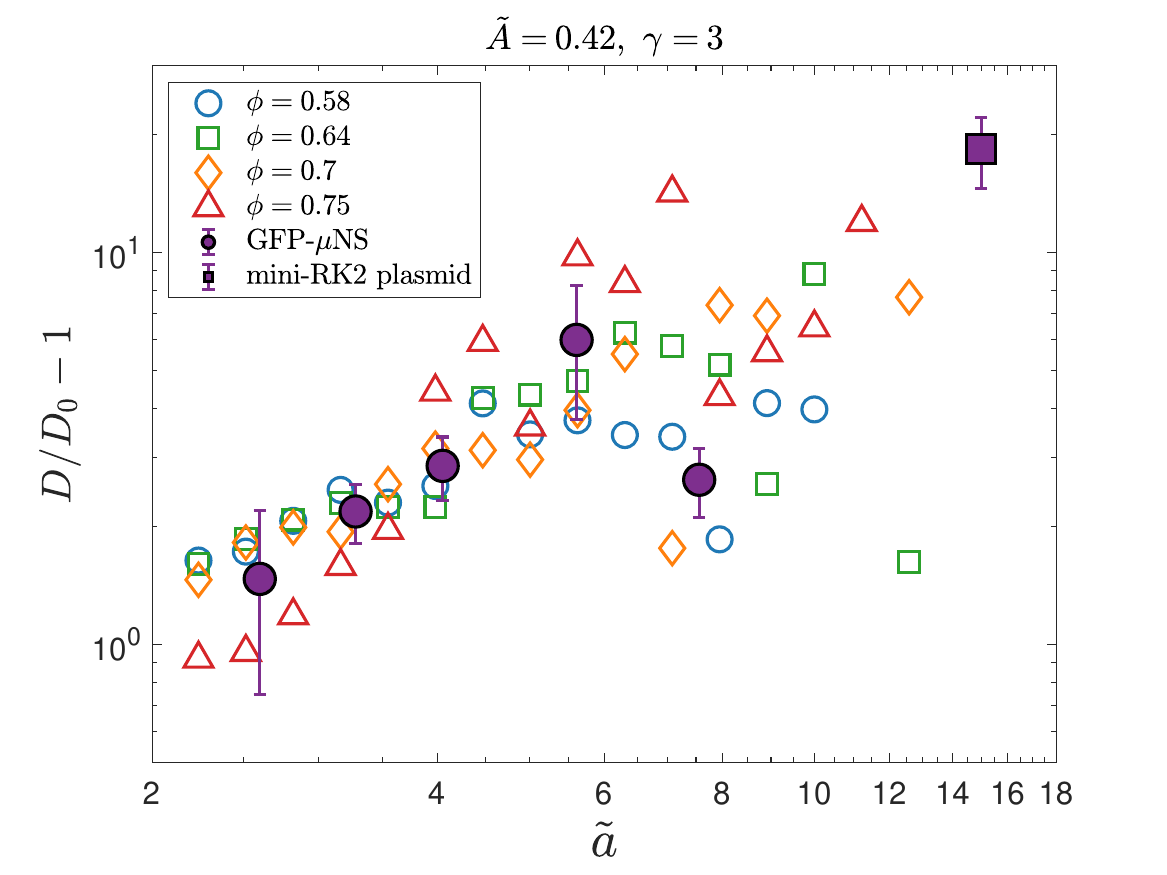}\\
	\captionof{figure}{The same analysis as Figure \ref{D_ratio_wExp}a in the main text, using a larger $\Delta \tilde{t}$ to compute the diffusion constant (see Appendix \ref{SuppleFig}---figure \ref{D_over_Ddilute_varDt}). The results are binned over particles with a bin interval of $0.05$ in the log$_{10}$ scale.}
	\label{D_ratio_wExp_varDt}
\end{center}

\begin{center}
	\includegraphics[width=0.6\textwidth]{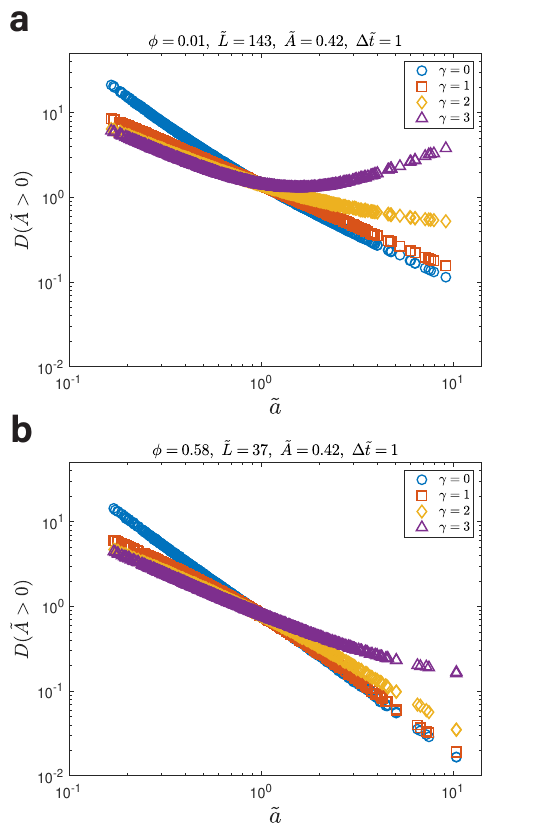}\\
	\captionof{figure}{The diffusion constants of active systems. (a) $\phi=0.01$ and (b) $\phi=0.58$. In the dilute limit, an active random force with $\gamma=3$ leads to a non-monotonic diffusion constant as a function of the particle radius. However, in systems with a high volume fraction, the diffusion constant decreases with the particle radius. In both figures, $\tilde{A}=0.42$ and the time interval to compute the diffusion constants is $\Delta \tilde{t}=1$. In both figures, $N=1000$. }
	\label{D_multi_phi}
\end{center}

\begin{center}
	\includegraphics[width=0.6\textwidth]{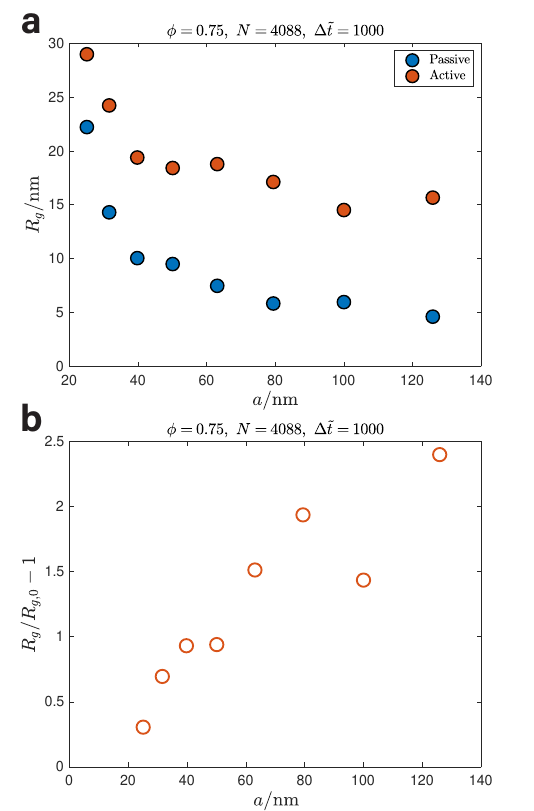}\\
	\captionof{figure}{Radii of gyration ($R_g$) for both passive and active systems with $\phi=0.75$. (a) Radii of gyration of large particles for both passive ($R_{g,0}$) and active systems ($R_g$) within a dimensionless time $1000$. The radii of gyration decrease linearly with the particle radius in both systems approximately. (b) The ratio of the radii of gyration between active and passive particles as $R_g/R_{g,0}-1$ increases linearly with the particle radius approximately, similar to experimental results. In both figures, the radius is converted to physical length using $a_0=10$ nm. The data are binned with a window of $0.1$ in the logarithmic space of radius. For the active system, $\tilde{A}=0.42$ and $\gamma=3$. In both figures, $N=4000$. }
	\label{Rg}
\end{center}

\begin{center}
	\includegraphics[width=1\textwidth]{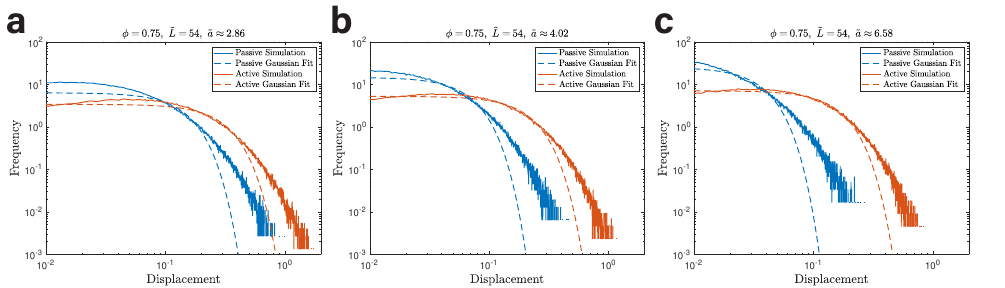}\\
	\captionof{figure}{Displacement distributions for passive and active systems with different particle radii. (a) The calculated distributions in both passive and active systems have a long tail compared with the Gaussian fits, which are the dashed lines. The displacement distributions are calculated over $50$ particles with similar radii within a time interval $\Delta \tilde{t}=1$. The average radius over these particles is $\tilde{a}\approx2.86$. (b) The same analysis as (a) but for $\tilde{a}\approx4.02$. (c) The same analysis as (a) but for $\tilde{a}\approx6.58$. For the active systems, $\tilde{A}=0.42$ and $\gamma=3$. In all figures, $\phi=0.75$ and $N=4000$.}
	\label{StepFreq}
\end{center}

\begin{center}
	\includegraphics[width=0.6\textwidth]{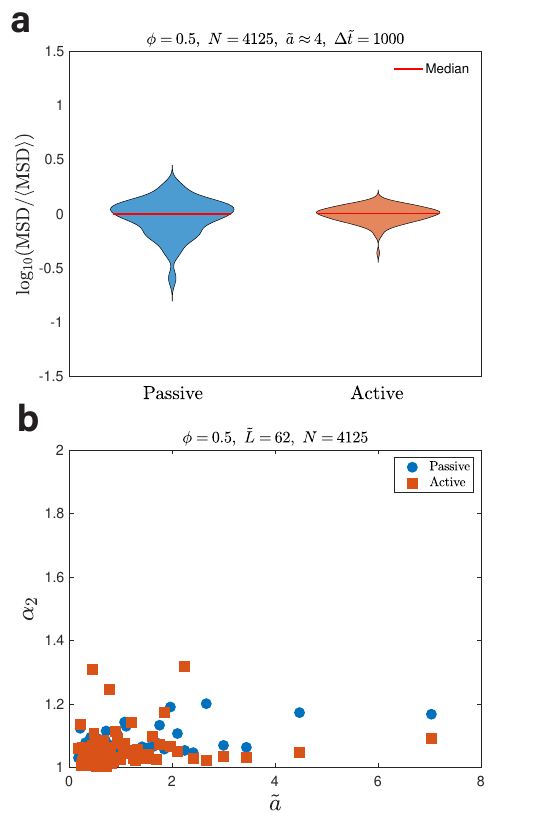}\\
	\captionof{figure}{Glassy-like properties are not obvious under low volume fraction. (a) With $\phi=0.5$, the distributions of MSDs for both passive and active particles whose $\tilde{a}\approx4$ become narrower, which means weaker dynamical heterogeneity. The MSDs within a dimensionless time 1000 are shown in the log scale and normalized by the average value. (b) With $\phi=0.5$, $\alpha_2$ of both passive and active particles are close to $1$, which indicates nearly Gaussian motions. The results are averaged with 50 particles in each bin. For the active systems, $\tilde{A}=0.42$ and $\gamma=3$. In both figures, $N=4000$. }
	\label{DHNGLow}
\end{center}

\begin{center}
	\includegraphics[width=0.6\textwidth]{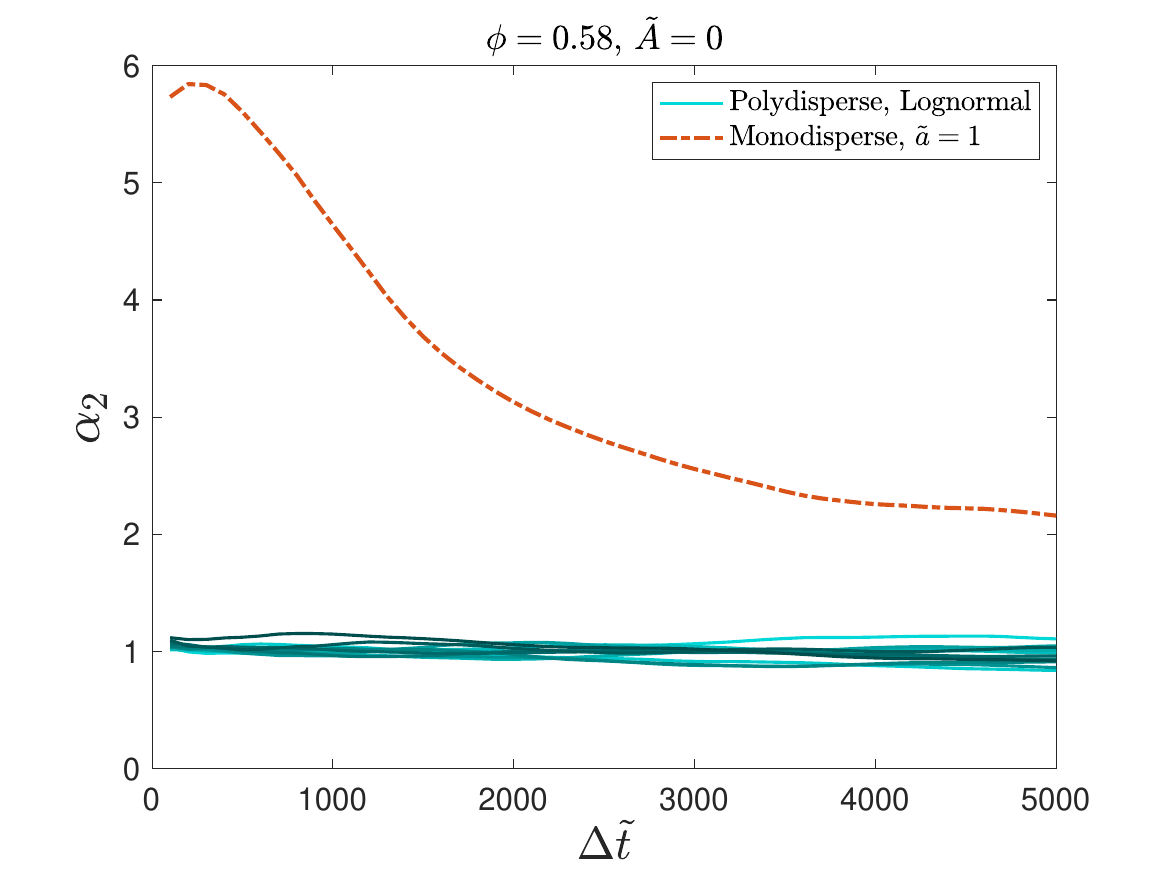}\\
	\captionof{figure}{The non-Gaussian parameter $\alpha_2$ for polydisperse and monodisperse systems. $\Delta \tilde{t}$ is the duration of the displacement. The green lines are particles with different radii from the polydisperse simulations. All green lines are close to $1$, which indicates Gaussian distribution. The orange line is obtained from the monodisperse system, indicating non-Gaussian displacement distributions. In both systems, $\phi=0.58$ and $\tilde{A}=0$. The $\alpha_2$ are calculated with displacements without the periodic boundary condition. In both systems, $N=1000$. }
	\label{alpha_PolyMono}
\end{center}

\begin{center}
	\includegraphics[width=0.6\textwidth]{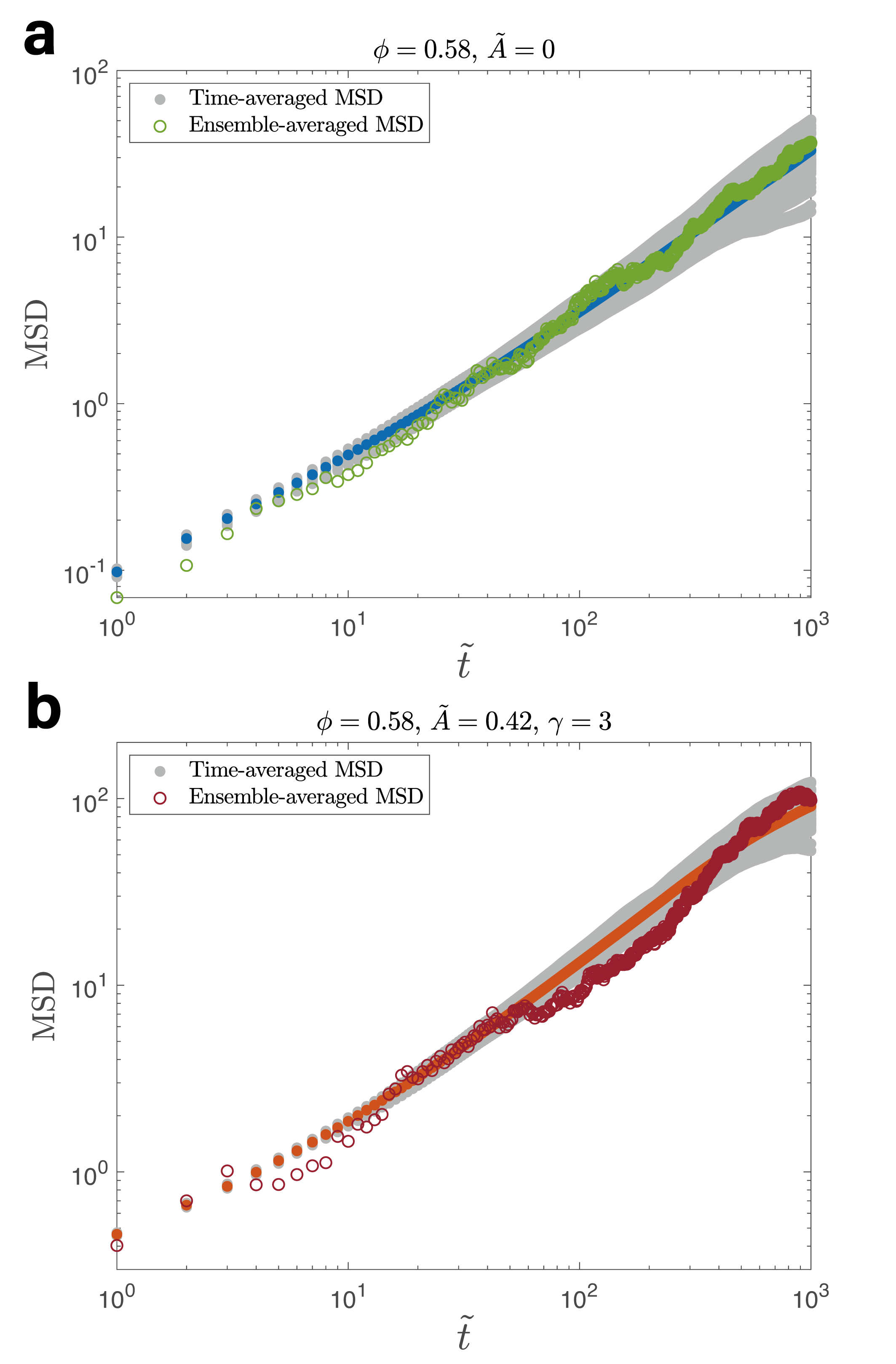}\\
	\captionof{figure}{Comparison between ensemble-averaged MSDs and time-averaged MSDs. (a) Analysis of passive systems. Ensemble-averaged MSDs are averaged over $100$ independent simulations (corresponding to distinct cells in experiments). Time-averaged MSDs are averaged over time using one simulation with total duration $\tilde{t}=10^4$. The blue circles are the averaged time-averaged MSD over independent simulations. (b) The same analysis of active systems where the orange circles are the averaged time-averaged MSD over independent simulations. In (a) and (b), $\phi=0.58$. For the active system, $\tilde{A}=0.42$ and $\gamma=3$. In both figures, $N=1000$.}
	\label{MSD_and_MSDtau}
\end{center}

\end{appendixbox}

\begin{appendixbox}
\label{friction}
\section{Effects of the active random force on the friction coefficient}
 Generally, the active random force can generate an additional friction coefficient $\zeta_i=Aa_i^\gamma/k_BT_{\text{act}}$ where $T_{\text{act}}$ is the active temperature \citep{Solon2022, Shakerpoor2021, Granek2022}. The equation of motion becomes
\begin{equation}
	(\eta_i+\zeta_i) \frac{\mathrm{d}r_{i,\alpha}}{\mathrm{d}t}=-\frac{\partial U}{\partial r_{i,\alpha}}+\xi_{i,\alpha}+\kappa_{i,\alpha}.  \label{rit_act}
\end{equation}
The active friction coefficient can be negligible if the active temperature $T_{\text{act}}$ is much higher than the thermal temperature $T$. In our model, the active random force comes from the collision of small background molecules, e.g., amino acids. The active temperature satisfies $k_BT_{\text{act}}=v^2 \tau \eta$, where $v$ is the velocity of the small background particle, $\tau$ is the relaxation time of the velocity's autocorrelation function, and $\eta=6\pi\nu a$ is the friction coefficient \citep{Solon2022}. For a passive system with $T_{\text{act}}=T$, $\tau=m/\eta$ where $m$ is the mass of amino acid. We take $m=110$ Da and $a=0.5$ nm for an average amino acid molecule \citep{Milo2015}, and find $\tau\approx 1.41\times10^{-15}$ s. The velocity can be estimated from the equipartition theorem $v=\sqrt{k_BT/m} = 1.50\times10^2$ m/s. Therefore, the characteristic ballistic length in a passive system $l=v\tau=2.12\times10^{-4}$ nm, which is extremely small. Due to active processes in living cells, active molecules are expected to have longer ballistic lengths, so the active temperature is expected to be higher than the thermal temperature \citep{gnesotto2018broken, Solon2022, Shakerpoor2021}. Assuming a ballistic length around $0.1$ nm, which is still much smaller than protein sizes, the active temperature is already much higher than the thermal temperature, $T_{\text{act}}=473T$. We confirm that the active friction coefficient is negligible given such a high ratio under the dilute limit (Appendix \ref{friction}---figure \ref{ActiveFriction}a). We also simulate Eq. (\ref{rit_act}) directly with  $T_{\text{act}}=100T$. The simulation results match the experimental data as the model without the active friction coefficient (Appendix \ref{friction}---figure \ref{ActiveFriction}b), further supporting the validity of our model.

\begin{center}
	\includegraphics[width=0.6\textwidth]{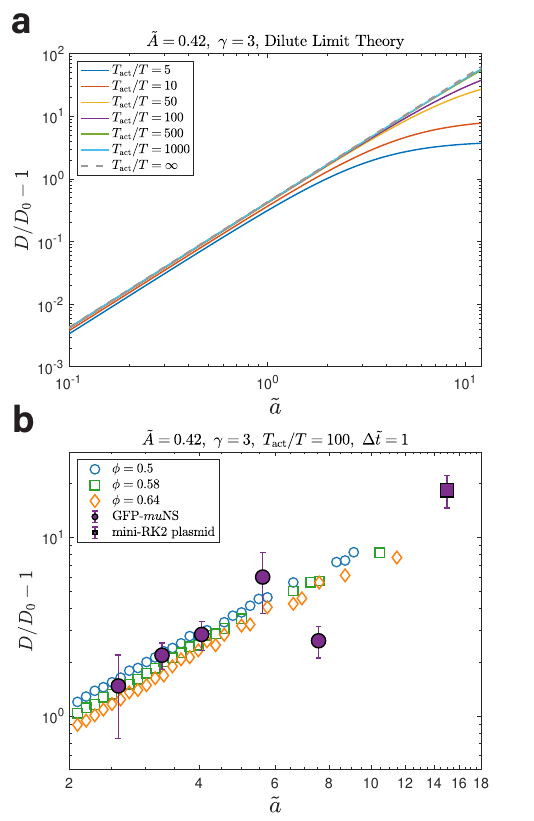}\\
	\captionof{figure}{Effects of the additional active friction coefficient. (a) Theoretical $D/D_0-1$ for different active temperatures $T_{\text{act}}$ from Eq. (\ref{rit_act}) under the dilute limit. When the active temperature is high, the effects of the active friction are negligible. (b) Comparison between simulations using Eq. (\ref{rit_act}) with $T_{\text{act}}=100T$ and experiments. The relative enhancements of diffusion constants from the experimental data still match the simulation results. In both figures, $\tilde{A}=0.42$ and $\gamma=3$. In (b), the results are binned over particles with a bin interval of $0.02$ in the log$_{10}$ scale and $N=1000$.}
	\label{ActiveFriction}
\end{center}
	
\end{appendixbox}

\end{document}